\documentclass[11pt]{article} 
\usepackage{hyperref} 
\pdfoutput=1

\title{Paramagnetic Leidenfrost Drops}
\author{Keyvan Piroird, Baptiste Darbois Texier, Christophe Clanet, David Qu\'er\'e\\ \\\vspace{6pt} {\small \emph{Physique et M\'ecanique des Milieux H\'et\'erog\`enes, CNRS, ESPCI, Paris France}}
 \\ {\small \emph{Ladhyx, CNRS, \'Ecole Polytechnique, Palaiseau, France} }}
\begin{document} 
\maketitle

\begin{abstract}
We present a \href{run:/anc/Paramagnetic_Leidenfrost_Drops_Small.mpg}{fluid dynamics video} showing the behavior of drops of liquid
oxygen, at room temperature. Due to their low boiling point, these drops
levitate on a cushion of their own vapour. This property gives them a high
mobility, as known more generally in such Leidenfrost situations.

But liquid oxygen is also paramagnetic, and thus likely to be manipulated
using a magnet. It is first shown that the shape of the drop can be
modified by changing the drop/magnet distance; approaching the magnet acts
as reinforcing gravity, so that the drops get flattened by this action.
The transformation is of course reversible: as the magnet is withdrawn,
the liquid recovers its quasi-spherical shape.

Magnets can also be used to trap the oxygen drops. As they pass above a
magnet, they slow down significantly, a consequence of their deformation:
despite a very low friction, the vibrations induced by the drop
deformations represent an important source of dissipation: below a
well-defined velocity, drops can even be stopped in the magnetic trap.
Quicker ones can be decelerated and then captured by series of magnets.
Oberving the same events from the top reveals the complex shapes adopted
by the liquid as it crosses the traps.

\end{abstract}

\end{document}